%% file: HllqqPaper.tex
\documentclass[3p,sort&compress]{elsarticle}
\usepackage[hyperindex,breaklinks]{hyperref}

\def\theabstract{%
A search for a heavy Standard Model Higgs~boson decaying via
$\htollqq$, where $\ell=e$ or $\mu$, is presented.  The search
uses a data set of $pp$ collisions at $\sqrt{s}=7\tev$,
corresponding to an integrated luminosity of \lumi\ collected in 2011
by the ATLAS detector at the CERN LHC.  No significant excess
of events above the estimated background is found.  Upper limits at
95\%~confidence level on the production cross section of a Higgs~boson
with a mass in the range between 200 and $600\gev$ are derived.
A Standard Model Higgs boson with a mass in the range
$300\gev \le \mH \le 322\gev$ or $353\gev \le \mH \le 410\gev$ is excluded at 95\%~CL.
The corresponding expected exclusion range is $351\gev \le \mH \le 404\gev$ at 95\%~CL.
}

\usepackage{preprintcover} 
 \PreprintCoverPaperTitle{\papertitle} 
 \PreprintIdNumber{CERN-PH-EP-2012-125} 
 \PreprintCoverAbstract{\theabstract} 
 \PreprintJournalName{Physics Letters B}

\usepackage{atlasphysics}
\usepackage{subfigure}
\usepackage{multirow}
\usepackage{amsmath}
\usepackage{amssymb}
\usepackage{lineno}
\usepackage{ifpdf}
\ifpdf\usepackage{grffile}\else\fi

\newcommand\htollqq{\ensuremath{H\to ZZ\to\ell^+\ell^- q \bar{q}}}

\newcommand\jets{\mathrm{jets}}

\newcommand\mll{\ensuremath{m_{\ell\ell}}}
\newcommand\mjj{\ensuremath{m_{jj}}}
\newcommand\mlljj{\ensuremath{m_{\ell\ell jj}}}

\def\progname#1{{\sc #1}}

\def\Figref#1{Figure~\ref{#1}} %
\def\Figsref#1{Figures~\ref{#1}} %
\def\figref#1{Fig.~\ref{#1}}   %
\def\figsref#1{Figs.~\ref{#1}}   %
\def\secref#1{Section~\ref{#1}}
\newcommand\lumi{\ensuremath{4.7~\ifb}}

\def\papertitle{Search for a Standard Model Higgs boson in the mass range $200$--$600\gev$ in the \htollqq\ decay channel
with the ATLAS Detector}

\begin{document}

\begin{frontmatter}
\title{\papertitle}
\author{The ATLAS Collaboration}
\address{}
\journal{Physics Letters B}
\begin{abstract} 
\theabstract
\end{abstract}
\begin{keyword}
Standard Model Higgs Boson, ATLAS
\end{keyword}
\end{frontmatter}

\section{Introduction}
In the Standard Model (SM), the as-yet-unobserved
Higgs~boson~\cite{Englert:1964et,Higgs:1964pj,Guralnik:1964eu} gives
mass to the weak vector bosons and other particles.
Direct searches performed at the CERN
Large Electron-Positron Collider (LEP) excluded at 95\% confidence level (CL) the production of a SM
Higgs boson with mass $\mH$ less than $114.4\gev$~\cite{Barate:2003sz}.  Searches at the Fermilab
Tevatron $p\bar{p}$ collider have excluded at 95\% CL the regions 100--$106\gev$ and 147--$179\gev$~\cite{TEVNPH:2012ab}.
At the ATLAS experiment at the LHC, the search was extended as far as
$600\gev$ using up to $4.9~\ifb$ of $\sqrt{s}=7\tev$ data recorded through 2011
(including an earlier version of this analysis with less data),
ruling out the production of a SM Higgs boson at 95\% CL in the regions 112.5--$115.5\gev$,
131--$237\gev$, and 251--$468\gev$~\cite{Collaboration:2012si}.  Corresponding results from
CMS~\cite{Chatrchyan:2012tx}, using $4.6$--$4.8~\ifb$ of $\sqrt{s}=7\:\tev$ data,
excluded at 95\% CL the region 127--$600\gev$.

If $\mH$ is larger than twice the $Z$~boson mass, $\mZ$, the Higgs~boson is expected to decay to two
on-shell $Z$~bosons with a large branching ratio. %
This Letter reports a search for a SM Higgs~boson in the mass range $200$--$600\gev$ decaying to a
pair of $Z$~bosons, where one $Z$~boson decays into two leptons and the other to two quarks: $\htollqq$
with $\ell\equiv e,\mu$.  
The analysis uses the full data set of $\lumi$ recorded by the ATLAS experiment in 2011.  Previous
results from the ATLAS Collaboration in this channel~\cite{Aad:2011ec,Collaboration:2012si}%
, using up to $2.05~\ifb$ of data, excluded a SM Higgs~boson production cross section between 1.2
and 12 times the SM cross section over this mass range.
The corresponding exclusions from the CMS collaboration with
$4.6\ \ifb$ of data are between 1.0 and 4 times the SM cross section
over the same mass range~\cite{Chatrchyan:2012sn}.

\section{Data and Monte Carlo samples}
\label{sec:samples}

The data used in this search were recorded by the ATLAS experiment during the 2011 LHC run with $pp$
collisions at $\sqrt{s}=7\tev$.  They correspond to an integrated luminosity of approximately \lumi\
after data quality selections to require that all systems used in this analysis were operational.
The data were collected using single-lepton triggers with a transverse momentum ($\pt$) threshold of
20 to $22\gev$ for electrons and $18\gev$ for muons, supplemented with a dielectron trigger with a
threshold of $12\gev$.
The resulting trigger
criteria are about 95\% efficient in the muon channel and close to 100\% efficient in the electron
channel, relative to the selection criteria described below. Collision events are selected by
requiring a reconstructed primary vertex with at least three associated tracks with $\pt>0.4\gev$.
The average number of collisions per bunch crossing in this data sample is about nine.

The \htollqq\ signal is modelled with the \progname{powheg} Monte
Carlo (MC) event generator~\cite{powheg4,powheg5},
which calculates separately the gluon and vector-boson fusion Higgs~boson production mechanisms up
to next-to-leading order (NLO).  Generated signal events are hadronised with
\progname{pythia}~\cite{pythia}, interfaced to
\progname{photos}~\cite{Golonka:2005pn} to model final-state radiation and
\progname{tauola}~\cite{Jadach:1993hs,Golonka:2003xt} to simulate $\tau$ decays. 
The parton distribution function (PDF) is \progname{MRSTMCal}~\cite{Sherstnev:2008dm}.
The Higgs boson
$\pt$ spectrum is reweighted to match
Ref.~\cite{deFlorian:2011xf},
which provides QCD corrections up to NLO and QCD soft-gluon resummations up to next-to-next-to-leading logarithms. 
The small contribution from $Z$~boson decay to $\tau$~leptons
is also included.%

The Higgs boson production cross sections and decay branching
ratios %
as well as their uncertainties, are taken from
Refs.~\cite{LHCHiggsCrossSectionWorkingGroup:2011ti,LHCHiggsCrossSectionWorkingGroup:2012vm}.  The
predicted cross sections for the gluon fusion processes are based on calculations to
next-to-next-to-leading order (NNLO) in
QCD~\cite{Djouadi:1991tka,Dawson:1990zj,Spira:1995rr,Harlander:2002wh,Anastasiou:2002yz,Ravindran:2003um},
and also include QCD soft-gluon resummations up to next-to-next-to-leading
logarithms~\cite{Catani:2003zt} and NLO electroweak (EW)
corrections~\cite{Aglietti:2004nj,Actis:2008ug}.
These results are compiled in Refs.~\cite{Anastasiou:2008tj,deFlorian:2009hc,Baglio:2010ae} and assume
factorisation between QCD and EW corrections.  The cross sections for the vector-boson fusion
processes are calculated with full NLO QCD and EW
corrections~\cite{Ciccolini:2007jr,Ciccolini:2007ec,Arnold:2008rz} and approximate NNLO QCD
corrections~\cite{Bolzoni:2010xr}.  The uncertainty in the production cross section due to the
choice of the QCD scale is $^{+12}_{-8}\%$ for the gluon fusion process and $\pm 1\%$ for the
vector-boson fusion
process~\cite{LHCHiggsCrossSectionWorkingGroup:2011ti,LHCHiggsCrossSectionWorkingGroup:2012vm}.  The
uncertainty in the production cross section due to uncertainties in the
PDFs and $\alpha_{s}$ is $\pm 8\%$ for the gluon-initiated process and $\pm 4\%$ for
quark-initiated processes~\cite{Botje:2011sn,Alekhin:2011sk,Lai:2010vv,Martin:2009iq,Ball:2011mu}.
The Higgs boson decay branching ratio~\cite{Djouadi:1997yw} to the four-fermion final state is predicted by
\progname{prophecy4f}~\cite{Bredenstein:2006rh,Bredenstein:2006ha}.
The combined production cross section and decay branching ratio for the $\htollqq$ channel ranges
from $140\pm20$~fb for $\mH=200\gev$ to $10\pm2$~fb for $\mH=600\gev$.

The cross section calculations do not take into account the width of the Higgs~boson,
which increases from $1.4\gev$ at $\mH=200\gev$ to $120\gev$ at $\mH=600\gev$, and
which is
implemented through a relativistic Breit-Wigner line shape applied at the event generator level.  It
has been
suggested~\cite{Seymour:1995qg,Passarino:2010qk,Anastasiou:2011pi,LHCHiggsCrossSectionWorkingGroup:2012vm}
that effects related to off-shell Higgs~boson production and interference with other SM processes
may become sizeable for the highest masses ($\mH>400\gev$) considered in this search.  Currently, in
the absence of a full calculation for the different production mechanisms, a conservative estimate
of the possible size of such effects is included as a signal normalisation systematic uncertainty
parameterised as a function of $\mH$ as $1.5\times\mH^3[\mathrm{TeV}]$, for
$\mH\ge300\gev$~\cite{LHCHiggsCrossSectionWorkingGroup:2012vm}.

The $Z+\text{light-jets}$ background is modelled with the \progname{alpgen}
generator~\cite{Mangano:2002ea} with the \progname{cteq6l1} PDF set~\cite{Pumplin:2002vw},
interfaced to \progname{herwig}~\cite{Corcella:2000bw} for parton showers and hadronisation, while
\progname{sherpa}~\cite{Gleisberg:2008ta} with the \progname{cteq6l1} PDF set is used for
$Z+\textrm{heavy-flavour}$ events.  Top~quark production, both $t\bar{t}$ and single-top, is
modelled using the \progname{mc@nlo} generator~\cite{Frixione:2003ei} with the \progname{ct10}
PDF set~\cite{Lai:2010vv}, interfaced to \progname{herwig} for parton showers and hadronisation.

The SM $ZZ$ process is an irreducible background for $H\ra ZZ$. The $q\bar{q}\to ZZ$ process (also
$WZ$) is modelled using \progname{herwig} with the \progname{MRSTMCal} PDF set, interfaced to \progname{photos} and
\progname{tauola}.  Alternative samples with \progname{pythia} and \progname{mc@nlo} are used for
systematics studies: \progname{herwig} and \progname{pythia} use only leading-order matrix elements,
but they can generate off-shell vector bosons, while \progname{mc@nlo} generates only on-shell
bosons.  The $q\bar{q}\to ZZ$ production cross section has been calculated up to NLO in
QCD~\cite{Campbell:1999ah}.  Due to the large gluon flux at the LHC, NNLO gluon pair quark-box
diagrams ($gg\to ZZ$) are significant and the $q\bar{q}$ cross section is increased by 6\% to account for this
additional contribution~\cite{Binoth:2008pr}.

Those simulations that use \progname{herwig} for hadronisation use \progname{jimmy}~\cite{jimmy} for
the modelling of the underlying event, while \progname{pythia} and \progname{sherpa} implement their
own underlying event model.

\section{Event selection}

The ATLAS detector~\cite{Aad:2008zzm} has a forward-backward symmetric cylindrical
geometry%
\footnote{ATLAS uses a right-handed coordinate system with its origin
at the nominal interaction point (IP) in the centre of the detector and
the $z$-axis coinciding with the axis of the beam pipe.
The $x$-axis points from the IP to the centre of the LHC ring,
and the $y$-axis points upward. Cylindrical coordinates ($r$,$\phi$)
are used in the transverse plane,  $\phi$ being the azimuthal angle
around the beam pipe. The pseudorapidity is defined in terms
of the polar angle $\theta$ as $\eta = - \ln \tan(\theta/2)$.
}%
.  An inner tracking detector %
immersed in a~2 Tesla axial magnetic field covers
$|\eta|<2.5$ with silicon detectors and straw tubes.  A liquid-argon electromagnetic
calorimeter %
is divided into barrel ($|\eta|<1.475$), endcap ($1.375<|\eta|<3.1$), and
forward ($3.1<|\eta|<4.9$) regions.  Hadronic calorimeters %
(using liquid argon or
scintillating tiles as active materials) surround the electromagnetic calorimeter and cover $|\eta|<4.9$.
A muon spectrometer %
measures the deflection of muon tracks in the field of
three large toroidal magnets and covers $|\eta|<2.7$.  A three-level trigger
system %
selects events to be recorded for offline analysis.

The offline selection starts with the reconstruction of either a $Z\ra ee$ or a $Z\ra\mu\mu$ lepton pair.
Electron and muon candidates must satisfy $\pt>20\gev$ and $|\eta|<2.5$, in addition to
standard ATLAS quality requirements~\cite{Aad:2011mk,Aad:2010yt,atlasmuon2011}, and must also be
isolated from surrounding tracks.  Electrons within $\Delta R \equiv
\sqrt{(\Delta\eta)^2+(\Delta\phi)^2} = 0.2$ of a muon are rejected.  The two muons in a pair are
required to have opposite charge, but this requirement is not imposed for electrons because larger
energy losses from showering in material in the inner tracking detector lead to higher charge
misidentification probabilities.
The invariant mass of the lepton pair $\mll$ must lie within the range 83--$99\gev$, and
events with any additional
selected electrons or muons are rejected to reduce background from $WZ$~production where both bosons
decay leptonically.

\htollqq\ decays contain a pair of jets from $Z\to q\bar{q}$ decay.  %
Events are thus required to contain at least two jets with $\pt>25\gev$ and $|\eta|<2.5$.
Jets are reconstructed with the anti-$k_t$ algorithm~\cite{AntiKt} with radius parameter $R=0.4$. 
They are calibrated using energy- and $\eta$-dependent correction factors based on MC
simulation and validated with data~\cite{Aad:2011he};
this calibration includes effects of energy from additional proton-proton interactions.
Jets within $\Delta R=0.4$ of an electron or in which less
than 75\% of the momentum of the associated tracks originates from the primary vertex are rejected.
The missing transverse momentum, with magnitude \met, is the (negative) vectorial sum of the
transverse momenta of all cells in the calorimeters with $|\eta|<4.9$, calibrated appropriately
based on their identification as electrons, photons, $\tau$~leptons, jets, or unassociated calorimeter cells,
and all selected muons in the event~\cite{met}. Calorimeter deposits associated with muons are
subtracted from \met\ to avoid double counting.  Since no high-$\pt$ neutrinos are present in the signal,
events are required to satisfy $\met<50\gev$, which primarily reduces the background from $\ttbar$
production.

Jets originating from $b$-quarks can be discriminated from other jets (``tagged'') based on the
relatively long lifetime 
of hadrons containing $b$-quarks.  This discrimination 
is important for this analysis because about 21\% of signal events contain $b$-jets from $Z\ra\bbbar$
decay, while $b$-jets are produced less often ($\sim 2$\%) in the
dominant $(Z\rightarrow\ell\ell)$+jets background.
A jet is tagged by taking the set of tracks associated with the jet and looking for either a secondary
vertex or for tracks that have a significant impact parameter with respect to the primary event
vertex~\cite{btagnote2011}.  This information is combined into a single discriminating variable
and a selection is applied that gives an efficiency of about 70\% (20\%) for
identifying true $b$-jets ($c$-jets)
with a light-quark jet rejection
of about~130~\cite{btageff2012,btagmistag2012}.  To optimise the expected
sensitivity, the analysis is divided into ``tagged'' and ``untagged'' subchannels, containing events
with exactly two and with fewer than two $b$-tags, respectively.  Events with more than two $b$-tags
($<3\%$ of the data sample with two $b$-tags) are rejected.

Events are required to have at least one candidate $Z\ra q\bar{q}$ decay with dijet invariant
mass $\mjj$ within 70--$105\gev$ in order to be consistent with a $Z$~boson decay.  This selection
is asymmetric around the $Z$~boson mass to account for non-Gaussian tails extending to lower
masses.  The jets forming a candidate must also be separated by $\Delta R > 0.7$,
excluding phase space regions poorly
modelled by MC simulation.  For untagged events, all pairs of jets formed from the three
highest-$\pt$ jets are considered.  All such pairs are retained with unit weight, leading to the
possibility of multiple candidates per event.  The fraction of untagged events with multiple pairs
retained is 13--16\% (2--5\%) for the low-$\mH$ (high-$\mH$) selection defined below.  For tagged
events, the two tagged jets are used to form the dijet invariant mass; their energies are scaled up
by 5\% to take into account the average energy scale difference between heavy- and light-quark jets.
The resulting dijet invariant mass distributions are well described by the MC simulation, as shown
in \figref{fig:mjj}.

\def\figmjj{%
   \centerline{%
\includegraphics[width=0.45\textwidth]{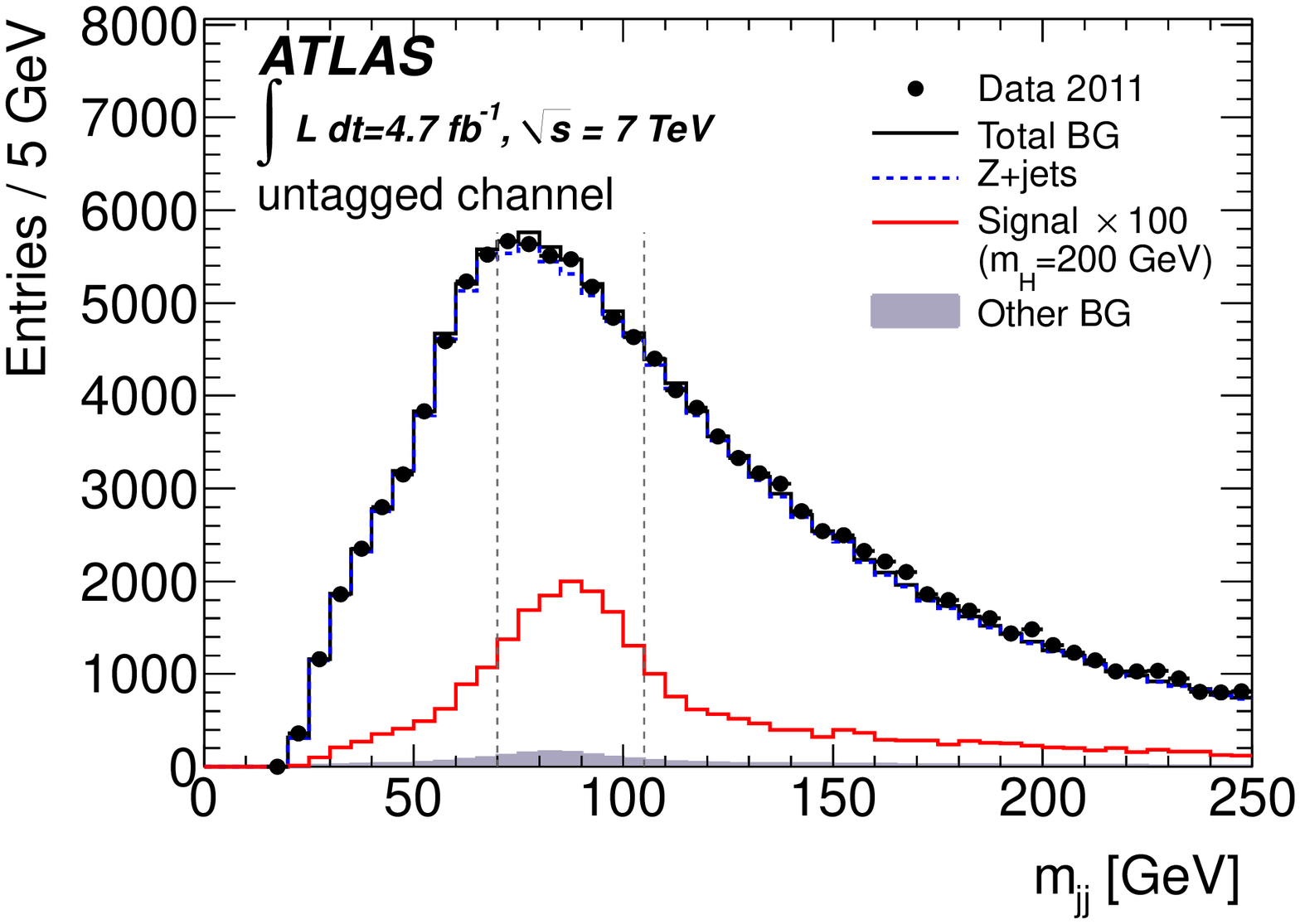}
\includegraphics[width=0.45\textwidth]{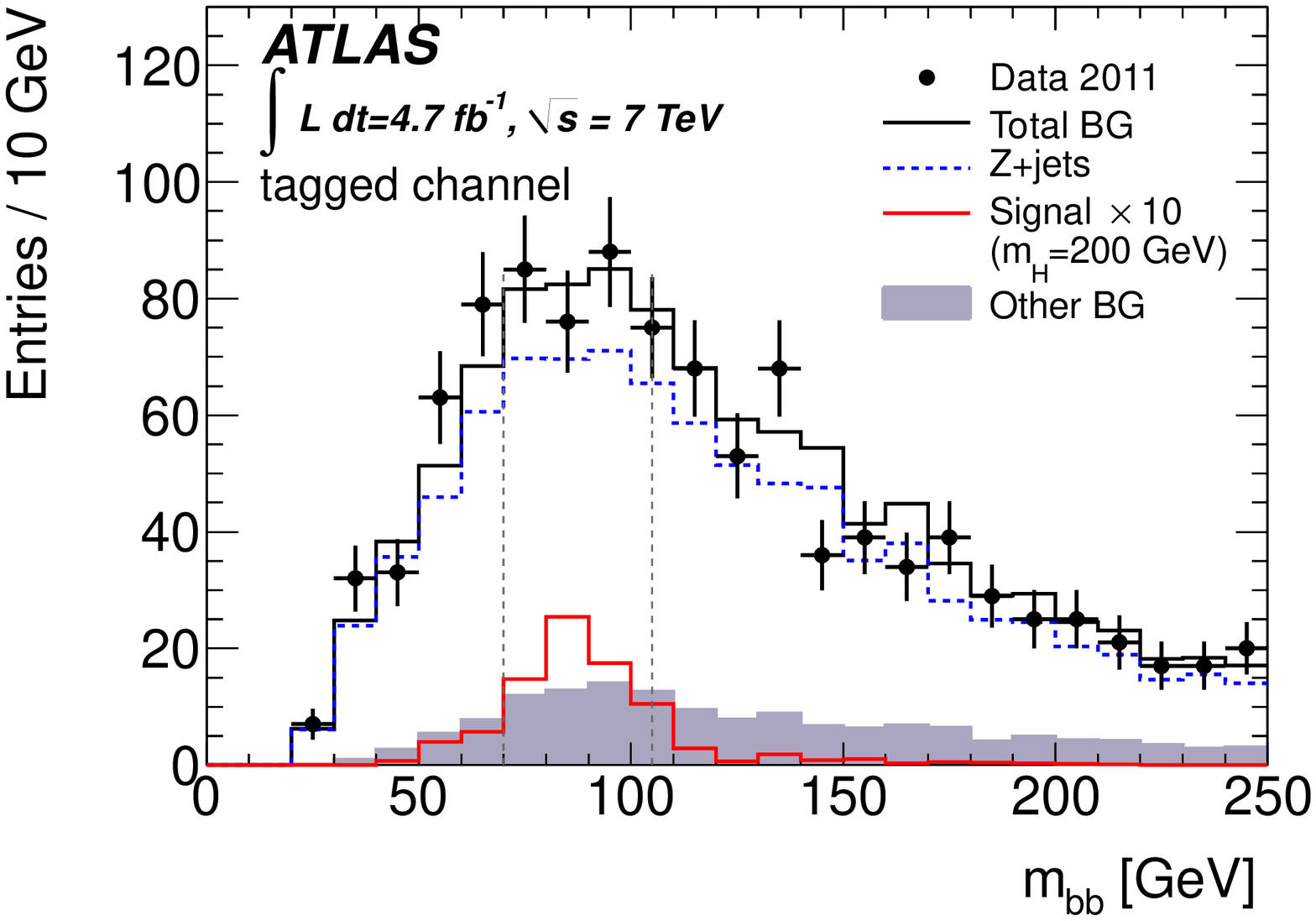}
    }
    \caption{Distributions of the invariant mass of selected dijet pairs, $\mjj$, for the data and
      the MC simulation, 
in the untagged (left) and tagged (right) samples, which contain fewer
      than two $b$-jets and exactly two $b$-jets respectively.    
      Scale factors and corrections described in Section~\ref{sec:background} have been applied.
      The signal has been scaled up by a factor of 100 (10) in the untagged (tagged) case to be
      more visible.  The vertical lines show the range of the $\mjj$ selection.
}%
}
\begin{figure}
  \figmjj
    \label{fig:mjj}
\end{figure}

The selection criteria above define the ``low-$\mH$'' selection, which is applied when searching
for a Higgs~boson with $\mH<300\gev$.  For higher Higgs~boson masses, the $Z$~bosons from the $H\to ZZ$
decay have large momenta in the laboratory reference frame, decreasing the opening angles between
their decay products.  Accordingly, in addition to the low-$\mH$ selection, the following
requirements are applied for $\mH\ge 300\gev$: the two jets must have $\pt>45\gev$ and the azimuthal
difference between the two leptons ($\Delta\phi_{\ell\ell}$) and the two jets
($\Delta\phi_{jj}$) must both be less than $\pi/2$.  These criteria define the ``high-$\mH$'' selection.

Following this event selection, a $\htollqq$ signal should appear as a peak in the invariant
mass distribution of the $\ell\ell jj$ system, with $\mlljj$ around $\mH$.  To improve the
Higgs~boson mass resolution, the energies of the jets forming each dijet pair are scaled
event-by-event
by a single
multiplicative factor to set the dijet invariant mass $\mjj$ to the nominal mass of the $Z$~boson.
The resolution is improved by a factor of 2.4 at $\mH=200\gev$; the improvement decreases with
increasing $\mH$ due to the increase in the natural width of the Higgs boson.
The total
efficiency for the selection of signal events is about $8\%$ over most of the mass range.

\section{Background estimates}
\label{sec:background}

The main background to this analysis is $Z$~boson production in association with jets ($Z+\jets$).
The shapes of the relevant kinematic distributions for this background are taken
from MC
simulation, with a small data-driven correction for the low-$\mH$ untagged selection, while the
normalisations for all selections are derived directly from data.

The flavour composition of the $Z$+jets sample is determined from three exclusive MC samples
containing at least one true $b$-jet, at least one true $c$-jet, and all light jets,
respectively.  The relative normalisations of the three components are adjusted by fitting the
distribution of the MC $b$-tagging discriminant to data.

To set the overall $Z+\jets$ normalisation, the $\mlljj$ distribution 
is compared between data and MC simulation for events in which the dijet invariant mass
$\mjj$ is in sidebands of the $Z$~boson mass: 40--$70\gev$ or 105--$150\gev$ (see
\figsref{fig:controlplots}(a) and~\ref{fig:controlplots}(b)).  The numbers of events in the sidebands, after subtraction of the
contribution from other background sources, are then used to derive scale factors to correct
the normalisation of the $Z+\jets$ MC simulation to that observed in the data.  The scale factors are
determined for the untagged channel separately for the low- and high-$\mH$ selections; the results
agree within statistical uncertainties.  In the tagged channel, there are too few events in the sidebands to
determine the scale factor for the high-$\mH$ selection, hence the low-$\mH$ scale factor is used
for both selections. Since the top~quark background is not negligible, the $Z+\jets$ MC normalisations are
determined in a simultaneous fit to the $Z+\jets$ control region and the corresponding top~quark control
region (see below).  The overall data to MC scale factors for $Z+\jets$ 
are approximately 0.9 for
light-jets, 1.9 for $c$-jets, and 1.5 for $b$-jets.

\def\figcontrolplots{%
   \centerline{%
\subfigure[$\mjj$ sidebands, untagged sample.]{\includegraphics[width=0.45\textwidth]{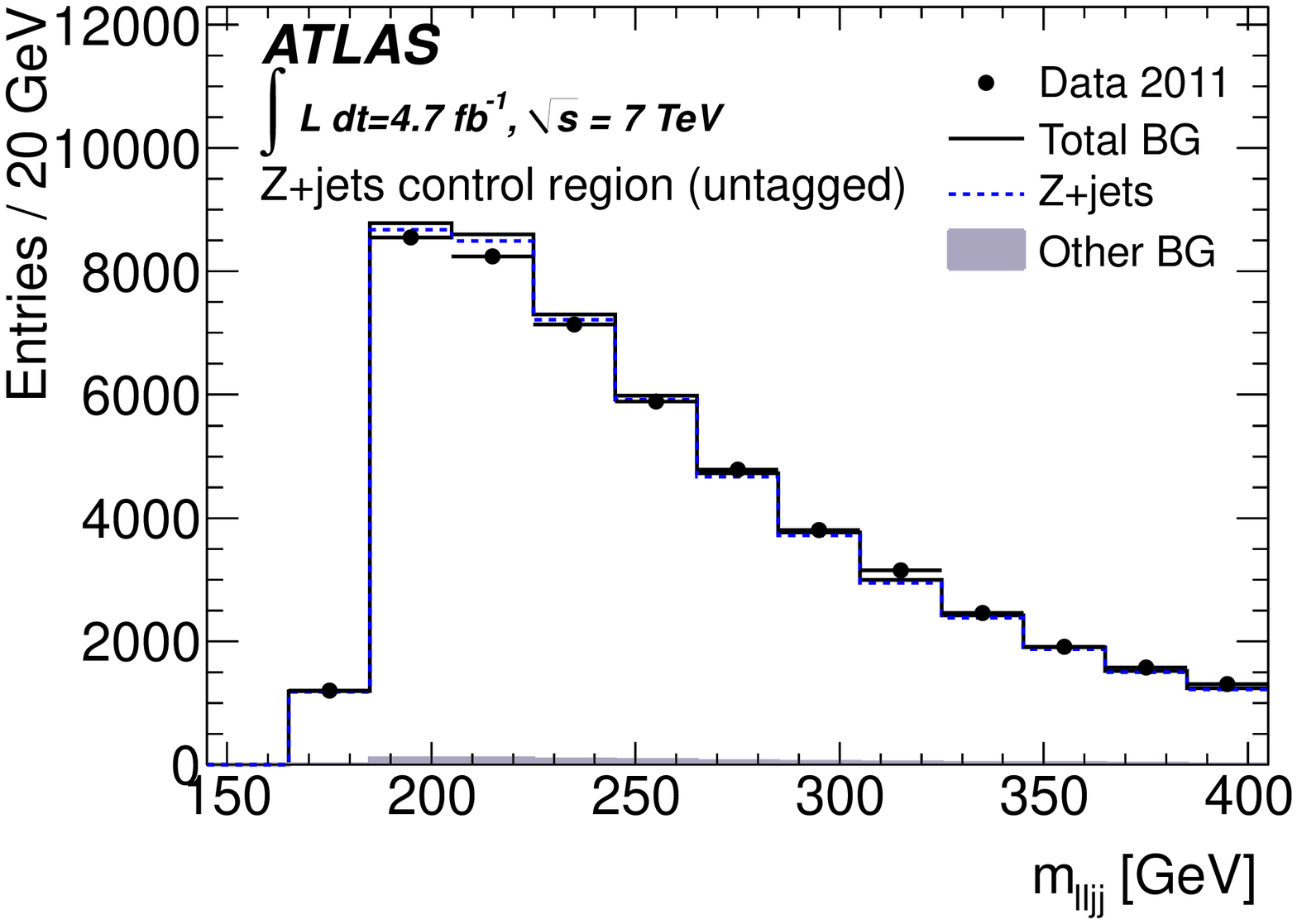}}
\subfigure[$\mjj$ sidebands, tagged sample.]{\includegraphics[width=0.45\textwidth]{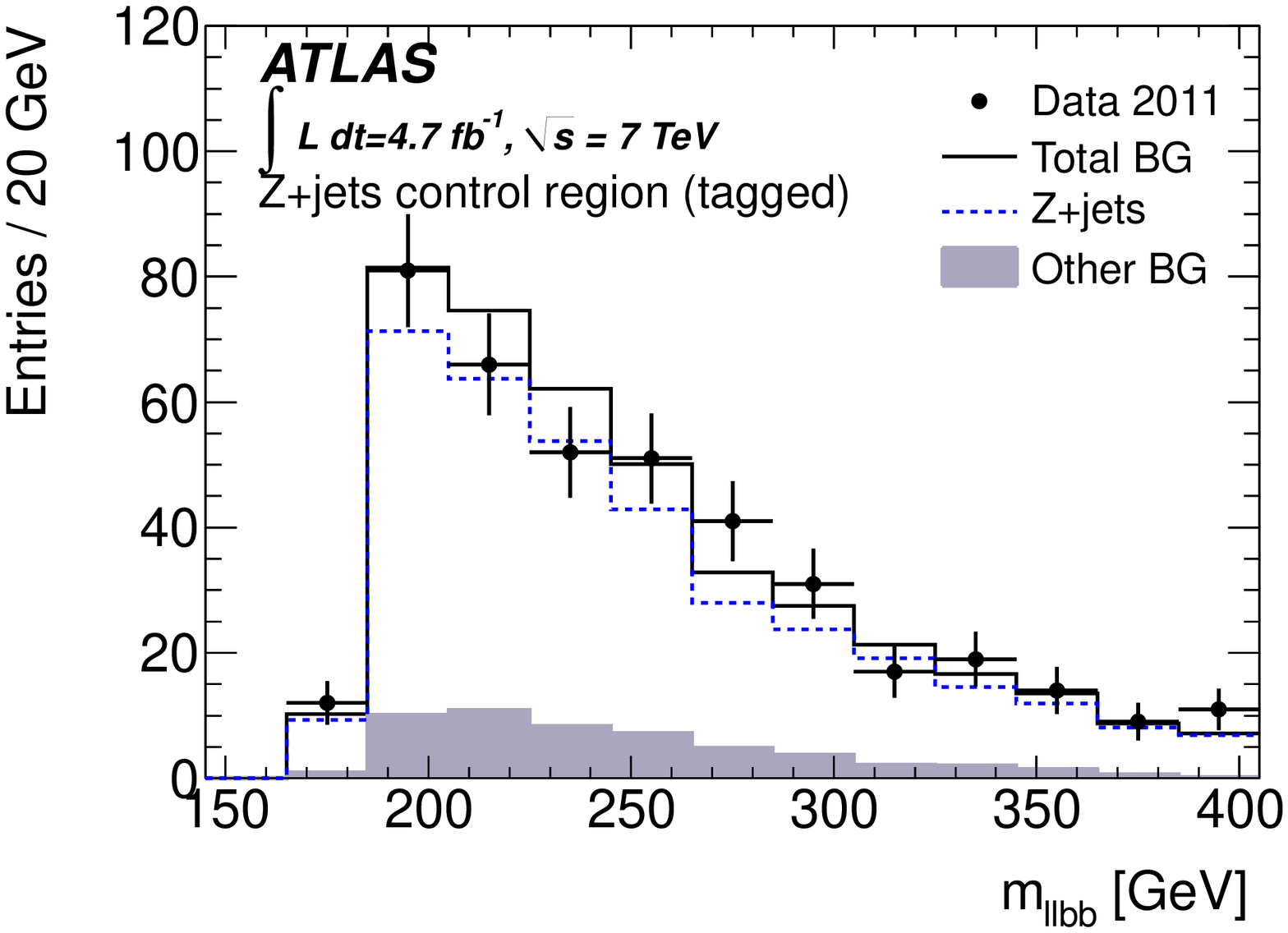}}
    }
   \centerline{%
\subfigure[$\mll$ sidebands,untagged sample.] {\includegraphics[width=0.45\textwidth]{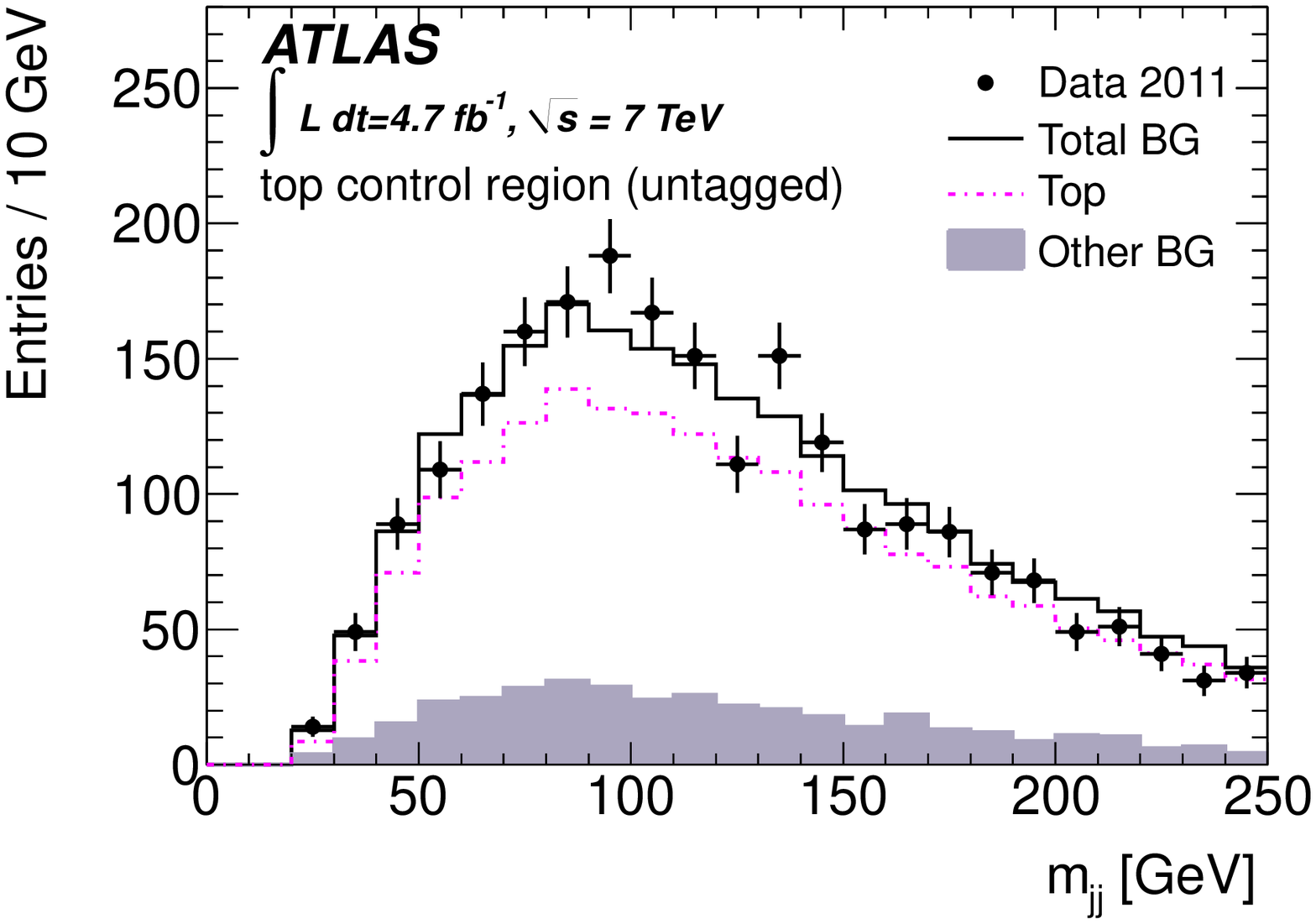}}
\subfigure[$\mll$ sidebands,tagged sample.] {\includegraphics[width=0.45\textwidth]{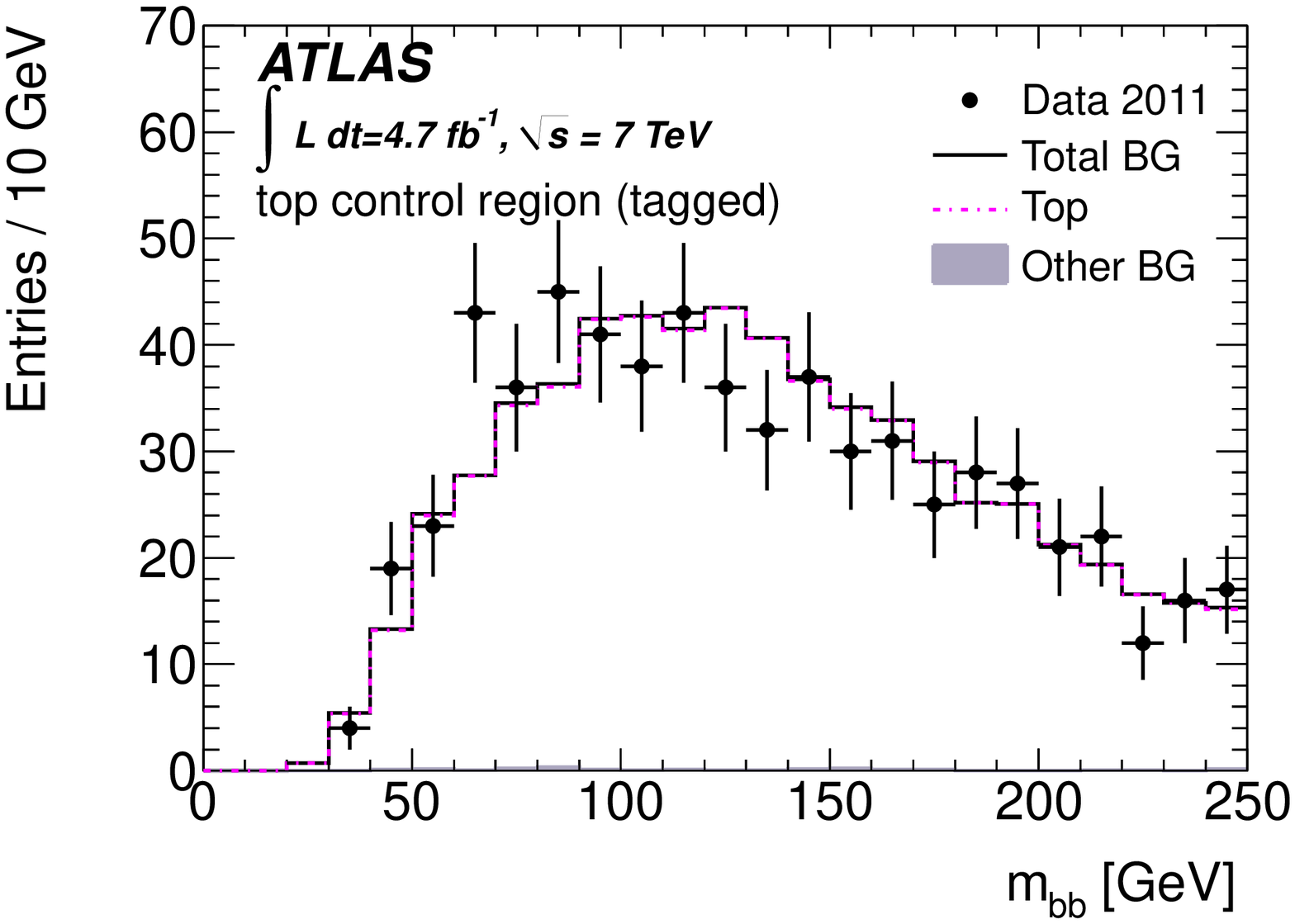}}
    }
  \caption{Distributions from the background control samples, after application
    of scale factors, for the low-$\mH$ selection.  Top row: the $\ell\ell jj$ invariant mass
    for $40\gev<\mjj<70\gev$ or $105\gev<\mjj<150\gev$
    for (a) the untagged
    and (b) the tagged sample.
    Bottom row: the invariant mass of the $jj$ system
    for events with $60\gev<\mll<76\gev$ or
    $106\gev<\mll<150\gev$  and $\met>50\gev$  for
    (c) the untagged sample 
    and (d) the tagged sample.}%
}
\begin{figure}
  \figcontrolplots
    \label{fig:controlplots}
\end{figure}
In the $\mjj$ sidebands of the untagged low-$\mH$ selection, the $Z+\jets$ MC simulation is about 3\% above the data
at  $\mlljj=200\gev$ and about 1\% below it at $\mlljj=300\gev$ (see \figref{fig:controlplots}(a)).  
Since similar results are seen for both the low and high mass sidebands, a linear fit to the ratio of data to MC simulation in the $\mlljj$ sideband
distribution is used to correct the prediction in the signal region.  
For the high-$\mH$ untagged selection and the tagged selections no difference between the data
and MC distributions is seen within statistical uncertainties.
Thus, no correction is applied to these samples, but similar fits to the one described above are
used to evaluate systematic uncertainties on the $Z+\jets$ $\mlljj$ shape.

The second most significant background is top~quark production, which is most important in the tagged
channel.  The shapes of the relevant kinematic distributions are taken from MC simulation and the
normalisation from data, using the $\mll$ sidebands, 60--$76\gev$ or 106--$150\gev$, with the $\met$
selection reversed.  \Figsref{fig:controlplots}(c) and \ref{fig:controlplots}(d) show the $\mjj$ distributions for these control
regions for the untagged and tagged selections respectively; good agreement is found after scaling up the
MC prediction by about 5\% for the untagged selection and 20\% for the tagged selection.

As in Ref.~\cite{Aad:2011ec}, the small irreducible background from diboson ($ZZ$ and $WZ$) production is estimated
directly from MC simulation. The background due to multijet events in which jets are
misidentified as isolated electrons is estimated from data using a sample of events containing
electron candidates that fail the selection requirements but pass loosened requirements.  The
multijet background to the muon channel was found to be negligible.  The background from $W+\jets$
production was also found to be negligible.

\section{Systematic uncertainties}

The theoretical uncertainties on the Higgs~boson production cross section from
Refs.~\cite{LHCHiggsCrossSectionWorkingGroup:2011ti,LHCHiggsCrossSectionWorkingGroup:2012vm} are
15--20\% for the gluon fusion process and 3--9\% for the vector-boson fusion process, depending on
the Higgs~boson mass.  As mentioned in \secref{sec:samples}, an additional uncertainty $\propto
\mH^3$ is applied for $\mH\ge300\gev$.  The selection efficiency uncertainty due to the production
process modelling is estimated by varying parameters of the signal MC simulation, including the
amount of initial- and final-state radiation, the factorisation and normalisation scales, and the
underlying event model; a further comparison uses \progname{pythia} instead of \progname{powheg}.
This procedure gives a 3\% (12\%) uncertainty for the low- (high-) $\mH$ selection.

The uncertainty in the normalisation of the $Z+\jets$ background from the procedure described in
\secref{sec:background} is evaluated by comparing the scale factors obtained from the upper or lower
sideband separately. The uncertainty is taken as the difference between the scale factors or the
statistical uncertainty, whichever is larger.  This procedure 
gives 1.7\% for the low-$\mH$ untagged
selection, 2.2\% for the high-$\mH$ untagged selection, and 5.5\% for both tagged selections.  The
uncertainty in the flavour composition of the $Z+\jets$ background is estimated by varying the
relative fraction of $Z+c$-jets by $\pm 30\%$ as determined by altering the selection criteria
applied in the fitting procedure described in \secref{sec:background}.  An uncertainty due
to the modelling of the $\mlljj$ shape as described in \secref{sec:background} is also applied.
Additional uncertainties on the shape of the $Z+\jets$ background are estimated by finding
variations of the MC $m_{jj}$ and $Z$~boson $\pt$ distributions that sufficiently cover any
differences between MC simulation and data in the $m_{jj}$ sidebands.

The uncertainty in the normalisation of the $\ttbar$ background is derived from the statistical
uncertainties on the normalisation scale factors.  It is found to be 2.7\% for the untagged
selection and 4.0\% for the tagged selection.
The diboson cross sections have a combined 5\% QCD scale and PDF
uncertainty~\cite{LHCHiggsCrossSectionWorkingGroup:2012vm}; adding an additional 10\% uncertainty,
corresponding to the maximum difference seen between \progname{mc@nlo} and $K$-factor scaled
\progname{pythia} results, yields an overall uncertainty of 11\% on the diboson background
normalisation.  A 50\% systematic uncertainty is assigned to the normalisation of the multijet
background in the electron channel by comparing the result of fitting the $\mll$ distribution before
and after the requirement of at least two jets. An overall 3.9\% uncertainty from the integrated
luminosity~\cite{Aad:2011dr,lumi2011} is added to the uncertainties on all MC processes that are not
normalised from data (i.e. excluding $Z+\jets$ and top~quark production), correlated across all
samples.

Contributions to systematic uncertainties also arise from detector effects, including the lepton and
jet trigger and identification efficiencies, the energy or momentum calibration and resolution of
the leptons and jets, and the $b$-tagging efficiency and mistag rates.  The dominant uncertainty on
the tagged sample comes from the $b$-tagging efficiency and corresponds to an average uncertainty
of 9\% on the signal~\cite{btageff2012,btagmistag2012}.  For the untagged sample, the uncertainties on the jet energy scale and
resolution contribute 3\% and 4\% respectively to the uncertainty
on the signal~\cite{Aad:2011he}.

The normalisations of the $Z+\jets$ and top~quark backgrounds are redetermined for each systematic
variation following the procedures described in~\secref{sec:background}.

\section{Results}

Table~\ref{tab:limits} shows the numbers of candidates observed in data for each
of the four selections compared with the background expectations.
\Figref{fig:results} shows the $\mlljj$ distributions
for both the tagged and untagged channels for the low- and high-$\mH$
selections.

\begin{table}[htb]
\caption{The expected numbers of signal and background candidates in the $\htollqq$ channel, along
  with the numbers of candidates observed in data, for an integrated luminosity of \lumi.  The
  low-$\mH$ analysis is applied when searching for a Higgs~boson with $\mH<300\gev$ and the
  high-$\mH$ analysis for $\mH \ge 300\gev$.  The first error indicates the statistical uncertainty, the second
  error the systematic uncertainty.}
\begingroup
\catcode`?=\active
\catcode`,=\active
\begin{center}
  {\small
    \def\cpm{$\,\pm\,$}
    \def\lcols#1{\multicolumn{3}{l}{#1}}
    \def\ccols#1{\multicolumn{3}{c}{#1}}
\newdimen\digitwidth
\newdimen\dotwidth
\setbox0=\hbox{\rm0}
\digitwidth=\wd0
\setbox0=\hbox{\rm.}
\dotwidth=\wd0
\def?{\kern\digitwidth}
\def,{\kern\dotwidth}
    \begin{tabular}{l*{4}{r@{\cpm}r@{\cpm}r}}
\hline
      & \multicolumn{6}{c}{Untagged}              & \multicolumn{6}{c}{Tagged}        \\
      & \ccols{Low-$\mH$}    & \ccols{High-$\mH$} & \ccols{Low-$\mH$}&\ccols{High-$\mH$}\\
\cline{2-13}%

$Z$+jets & 36190 & 80 & 640 &  1450,? & 14,? & 35,? & 239,? & 6,? & 15,? & 11,? & 1,? & 2,? \\
Top      &    85 &  3 &  10 &     7.1 &  0.7 &  0.8 &  23,? & 1,? &  3,? &  1.9 & 0.4 & 0.5 \\ 
Multijet &    15 &  0 &   8 &     0.2 &  0.0 &  0.1 &   \ccols{$<0.1$}   &  \ccols{$<0.1$}  \\
$ZZ$     &   348 &  3 &  47 &    25,? &  1,? &  3,? &  22,? & 1,? &  4,? &  2.3 & 0.3 & 0.4 \\ 
$WZ$     &   434 &  4 &  70 &    45,? &  1,? &  7,? &   0.7 & 0.2 &  0.3 &  \ccols{$<0.2$}  \\
\hline
Total background
         & 37070 & 80 & 670 &  1530,? & 14,? & 37,? & 285,? & 6,? & 18,? & 15,? & 1,? & 2,? \\
Data     &  \lcols{36898}   &  \lcols{1444}         &  \lcols{286}       & \lcols{18}       \\
\hline
\multicolumn{1}{c}{Signal}\\
\cline{1--1}%
$\mH=200\gev$
         &   118 &  2 &  19 &\lcols{}                  &  6.4 & 0.4 &  1.3 &\lcols{}            \\
$\mH=300\gev$
         &\lcols{}                   &    24.3 &  0.7 &  4.1 &\lcols{}              & 2.1 & 0.2 & 0.4 \\
$\mH=400\gev$
         &\lcols{}                   &    40.5 &  0.5 &  6.4 &\lcols{}              & 4.4 & 0.2 & 1.0 \\
$\mH=500\gev$
         &\lcols{}                   &    18.5 &  0.2 &  3.1 &\lcols{}              & 2.0 & 0.1 & 0.5 \\
$\mH=600\gev$
         &\lcols{}                   &     6.3 &  0.1 &  1.1 &\lcols{}              & 0.7 & 0.0 & 0.2 \\

\hline
    \end{tabular}
  }%
\end{center}%
\endgroup
\label{tab:limits}
\end{table}

\def\figresults{%
  \begin{center}
    \subfigure[Low-$\mH$, untagged selection.]
{\includegraphics[width=0.45\textwidth]      {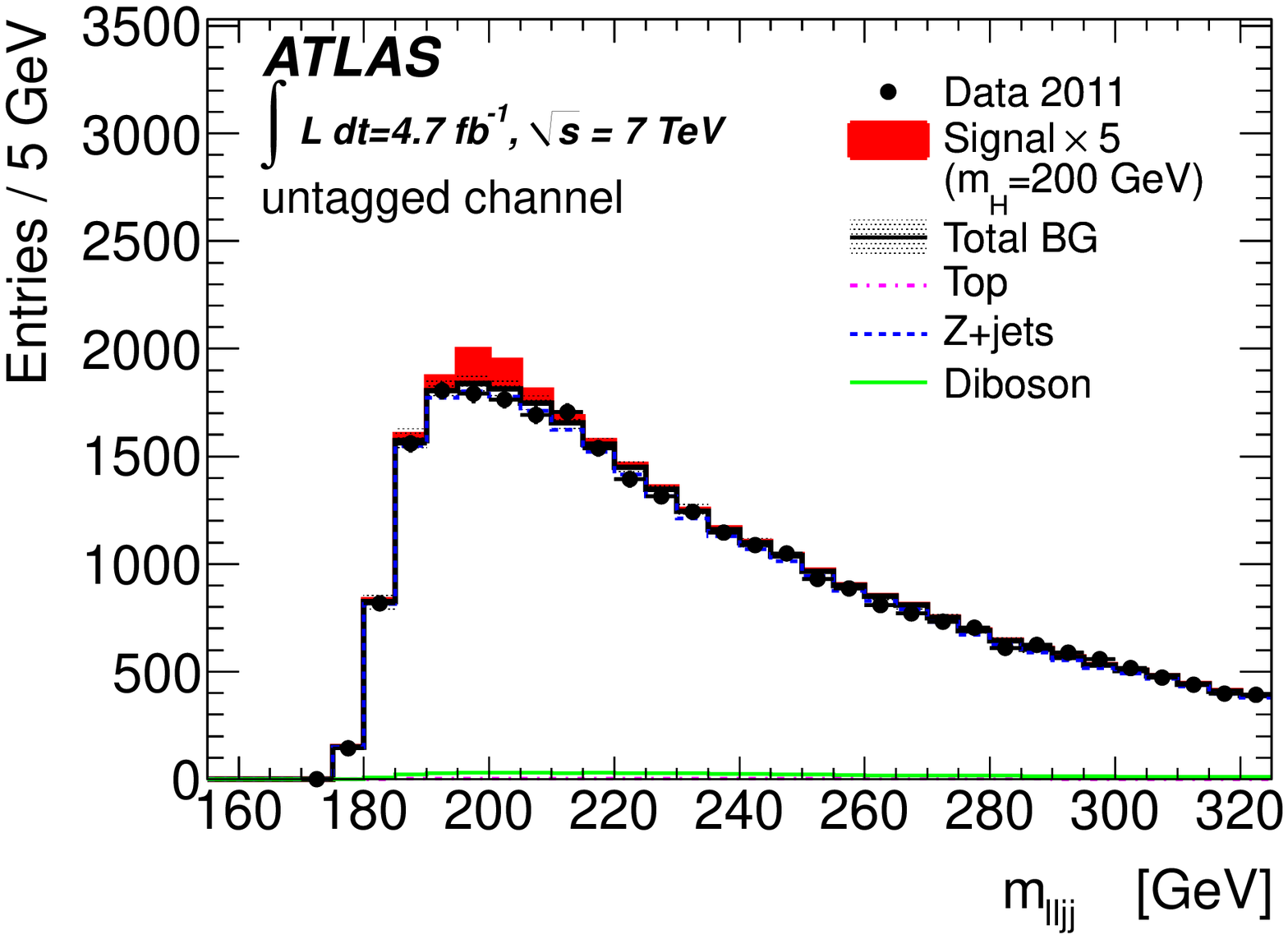}}
    \subfigure[Low-$\mH$, tagged selection.]
{\includegraphics[width=0.45\textwidth]      {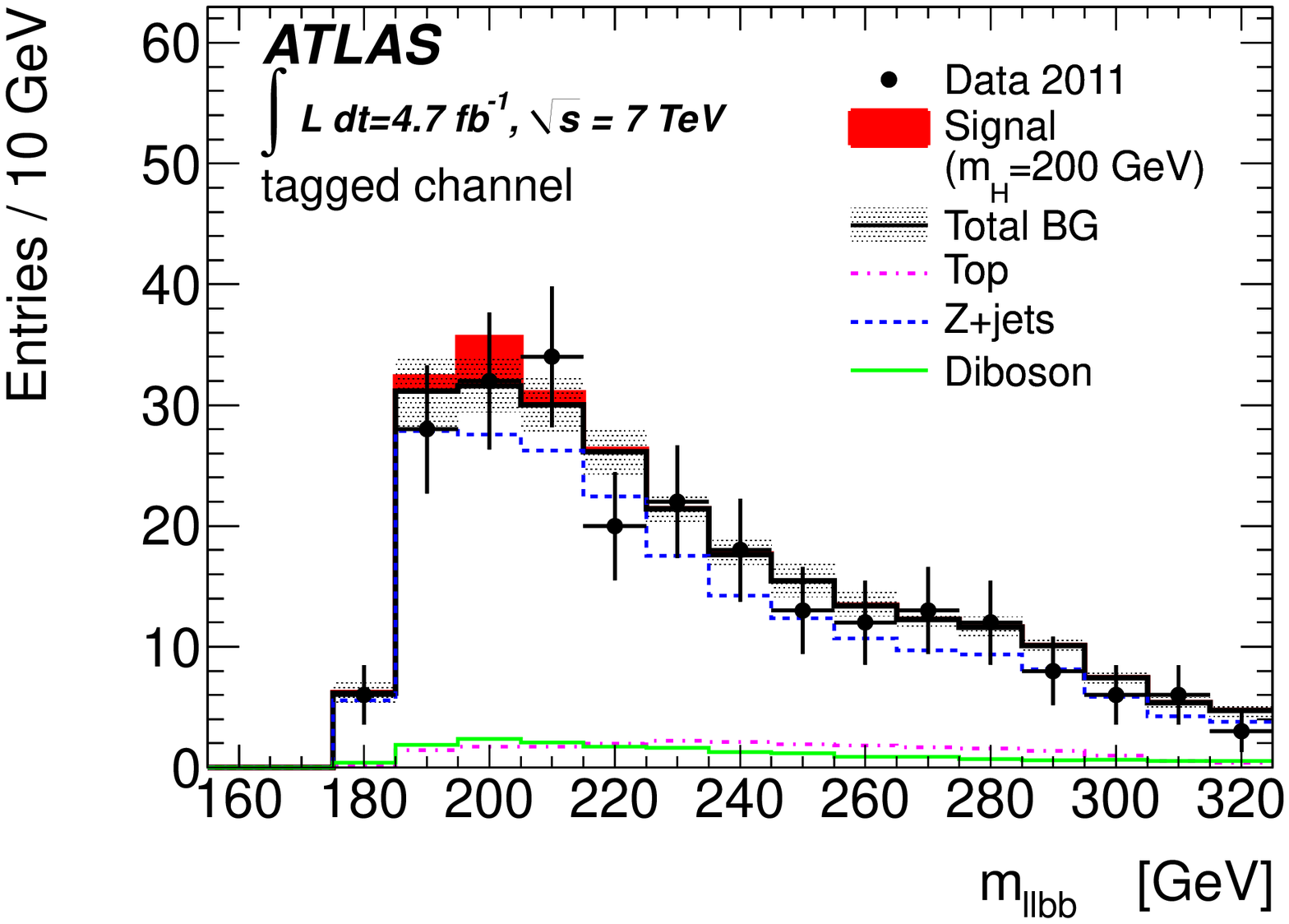}}
    \subfigure[High-$\mH$, untagged selection.]
{\includegraphics[width=0.45\textwidth]      {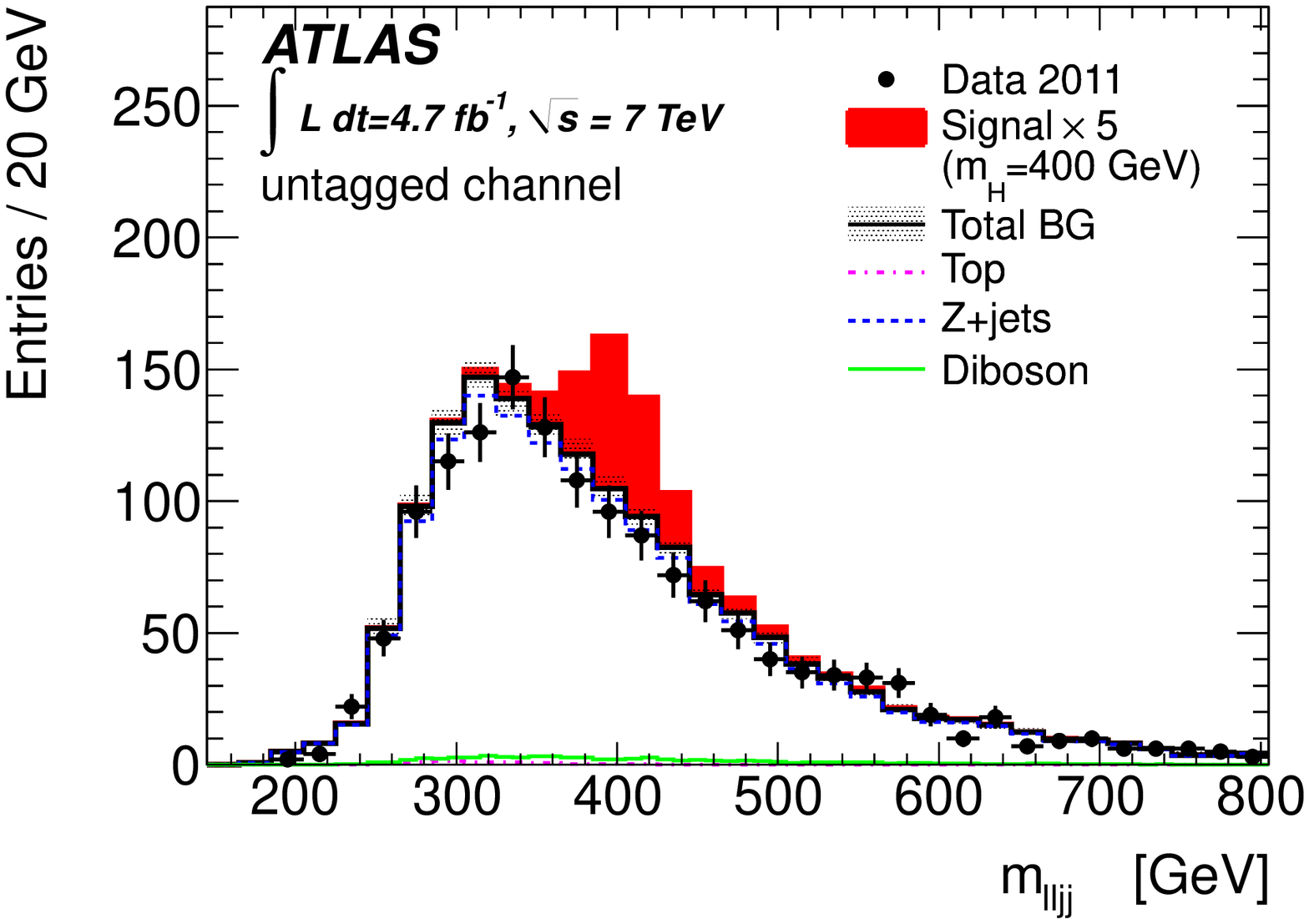}}
   \subfigure[High-$\mH$, tagged selection.]
{\includegraphics[width=0.45\textwidth]      {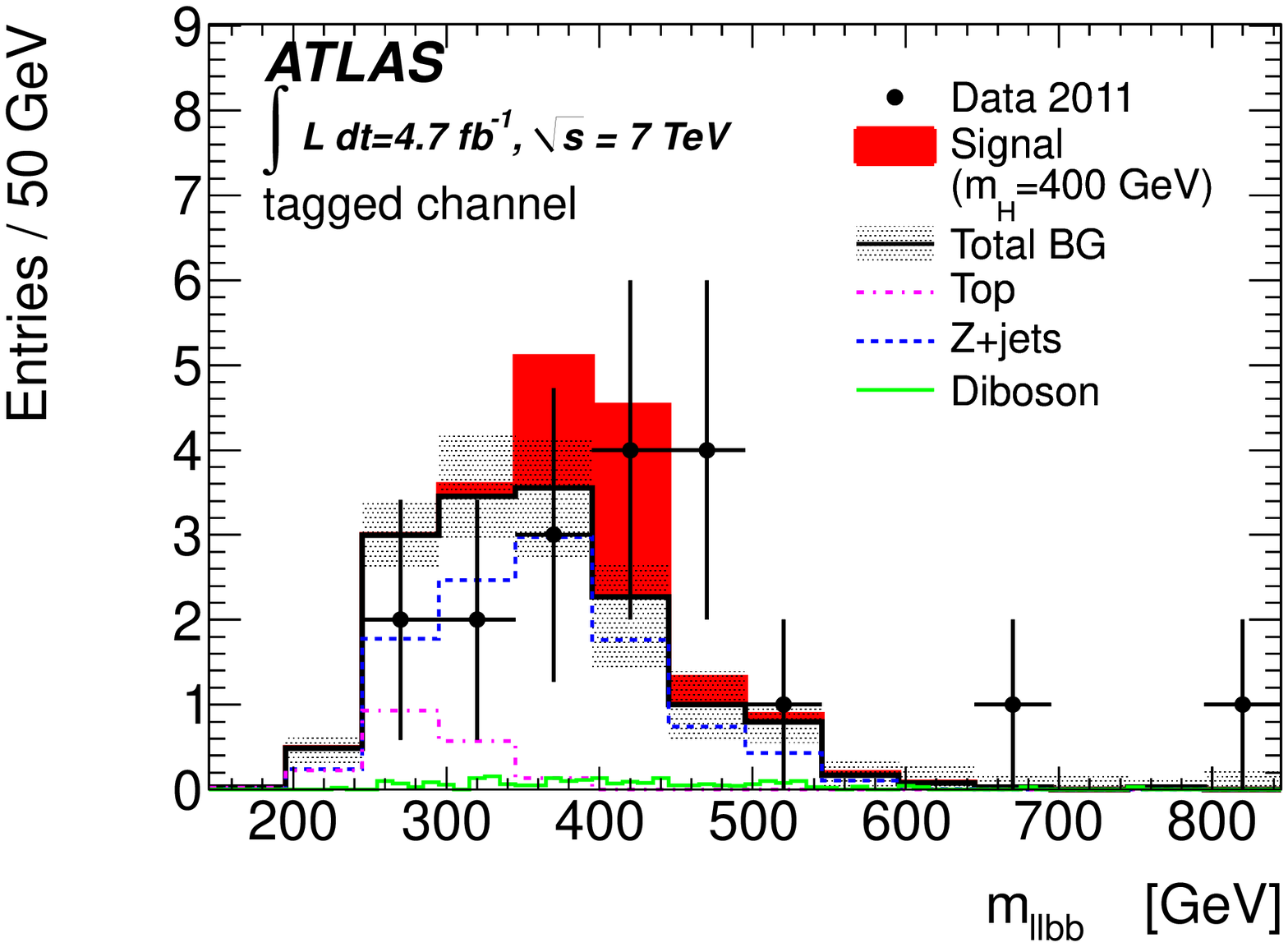}}
  \end{center}
  \caption{The invariant mass of the $\ell\ell jj$ system for both the untagged (a, c) and tagged
    (b, d) channels, for the low-$\mH$ (top row) and high-$\mH$ (bottom row) selections.  The hatched
    band represents the systematic error on the total background prediction. Examples
    of the expected Higgs~boson signal for $\mH=200$ and $400\gev$ are also shown; in the untagged
    plots (a, c), the signal has been scaled up by a factor of five to make it more visible.}%
}
\begin{figure}
  \figresults
  \label{fig:results}
\end{figure}

No significant excess of events above the expected background is seen; the smallest $p_0$ value
is 0.15 at $\mH=540$, where $p_0$ represents the probability that a background-only experiment would
yield a result that is more signal-like than the observed result.
Upper limits are set on the SM Higgs~boson cross section at 95\% CL as a
function of mass, using the $CL_s$ modified frequentist formalism with the profile likelihood test
statistic~\cite{Cowan:2010js,Read:2002hq}.  This method is based on a likelihood that compares, bin-by-bin
using Poisson statistics, the observed $\mlljj$ distribution to either the expected background or
the sum of the expected background and a mass-dependent hypothesised signal.  The tagged and
untagged channels, which contribute approximately equally across the $\mH$ range, are combined by
forming the product of their likelihoods; systematic uncertainties, with their correlations, are
incorporated as nuisance parameters.
\Figref{fig:hlimit} shows the resulting upper limit on the cross section for Higgs~boson production
and decay in the channel $\htollqq$ relative to the Standard Model cross section as a function
of the hypothetical Higgs~boson mass.

\def\fighlimit{%
  \centerline{%
\includegraphics[width=0.7\textwidth]{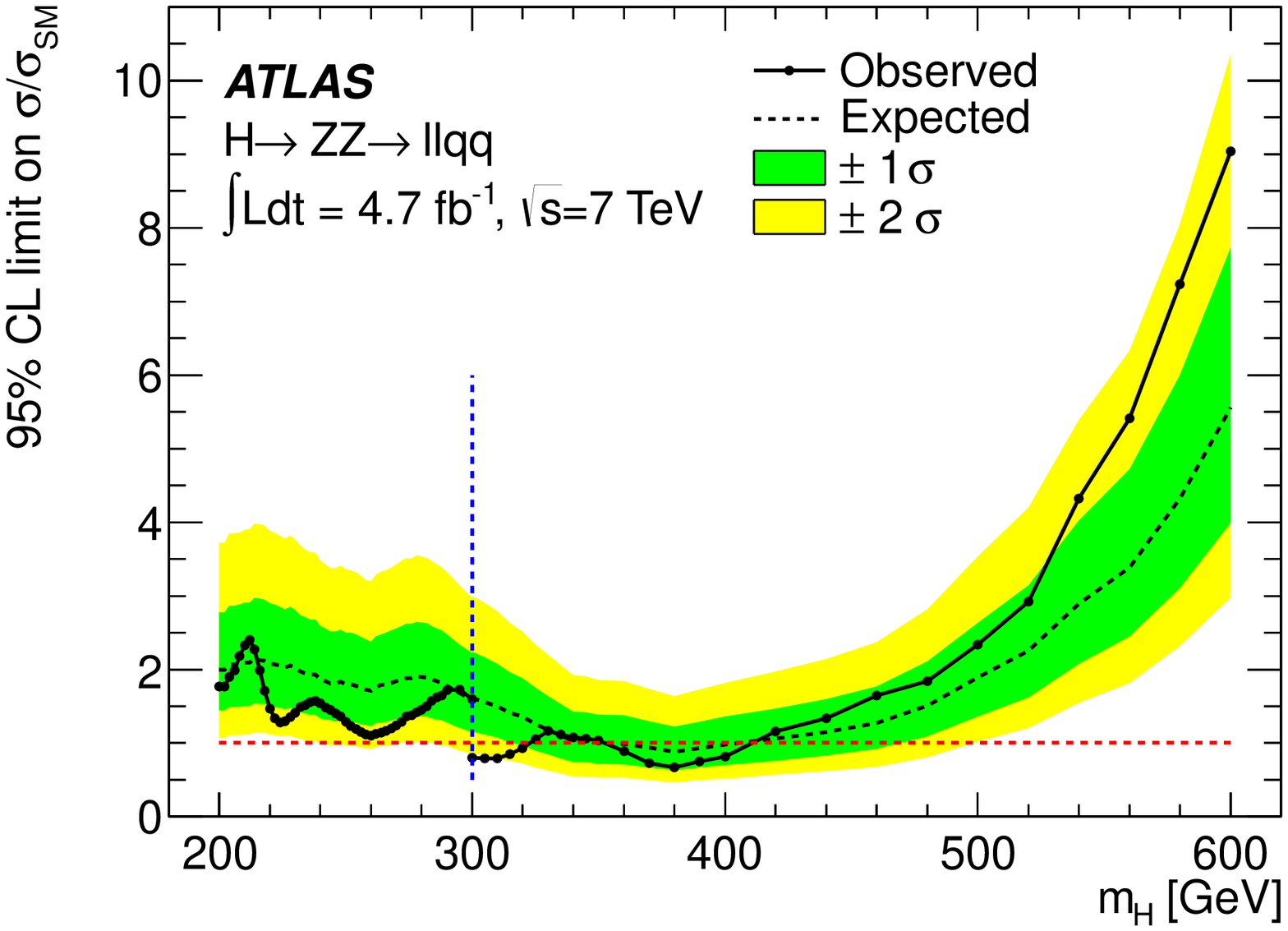}
}%
\caption{The expected (dashed line) and observed (solid line) upper limits on the total cross
  section divided by the expected SM Higgs~boson cross section, calculated using $CL_s$ at 95\%.
  The inner and outer bands, obtained from pseudo-experiments, indicate the $\pm 1\sigma$ and $\pm 2\sigma$ ranges in which the limit is expected to lie in the absence of a signal.  The horizontal
  dotted line shows the SM value of unity. The discontinuity in the limit at $\mH = 300\gev$ is due
  to the transition between the use of the low- and high-$\mH$ selections.}%
}
\begin{figure}[htb]
  \fighlimit
\label{fig:hlimit}
\end{figure}

\section{Summary}

A search for the SM Higgs~boson in the decay mode
$\htollqq$ has been performed in the Higgs mass range 200 to $600\gev$
using \lumi\ of $\sqrt{s}=7\tev$~$pp$ data recorded by the ATLAS experiment at
the LHC.  No significant excess over the expected background is found. 
A Standard Model Higgs~boson is excluded at a 95\% CL within the range
 $300\gev \le \mH \le 322\gev$ and $353\gev \le \mH \le 410\gev$.
The corresponding expected exclusion range is $351\gev \le \mH \le 404\gev$ at 95\% CL.

\section{Acknowledgements}

We thank CERN for the very successful operation of the LHC, as well as the
support staff from our institutions without whom ATLAS could not be
operated efficiently.

We acknowledge the support of ANPCyT, Argentina; YerPhI, Armenia; ARC,
Australia; BMWF, Austria; ANAS, Azerbaijan; SSTC, Belarus; CNPq and FAPESP,
Brazil; NSERC, NRC and CFI, Canada; CERN; CONICYT, Chile; CAS, MOST and NSFC,
China; COLCIENCIAS, Colombia; MSMT CR, MPO CR and VSC CR, Czech Republic;
DNRF, DNSRC and Lundbeck Foundation, Denmark; EPLANET and ERC, European Union;
IN2P3-CNRS, CEA-DSM/IRFU, France; GNAS, Georgia; BMBF, DFG, HGF, MPG and AvH
Foundation, Germany; GSRT, Greece; ISF, MINERVA, GIF, DIP and Benoziyo Center,
Israel; INFN, Italy; MEXT and JSPS, Japan; CNRST, Morocco; FOM and NWO,
Netherlands; RCN, Norway; MNiSW, Poland; GRICES and FCT, Portugal; MERYS
(MECTS), Romania; MES of Russia and ROSATOM, Russian Federation; JINR; MSTD,
Serbia; MSSR, Slovakia; ARRS and MVZT, Slovenia; DST/NRF, South Africa;
MICINN, Spain; SRC and Wallenberg Foundation, Sweden; SER, SNSF and Cantons of
Bern and Geneva, Switzerland; NSC, Taiwan; TAEK, Turkey; STFC, the Royal
Society and Leverhulme Trust, United Kingdom; DOE and NSF, United States of
America.

The crucial computing support from all WLCG partners is acknowledged
gratefully, in particular from CERN and the ATLAS Tier-1 facilities at
TRIUMF (Canada), NDGF (Denmark, Norway, Sweden), CC-IN2P3 (France),
KIT/GridKA (Germany), INFN-CNAF (Italy), NL-T1 (Netherlands), PIC (Spain),
ASGC (Taiwan), RAL (UK) and BNL (USA) and in the Tier-2 facilities
worldwide.

\bibliographystyle{elsarticle-num}
\bibliography{HllqqPaper}

\clearpage
\input{atlas_authlist}
\end{document}
\bye

%% file: atlas_authlist.tex
\begin{flushleft}
{\Large The ATLAS Collaboration}

\bigskip

G.~Aad$^{\rm 48}$,
B.~Abbott$^{\rm 111}$,
J.~Abdallah$^{\rm 11}$,
S.~Abdel~Khalek$^{\rm 115}$,
A.A.~Abdelalim$^{\rm 49}$,
O.~Abdinov$^{\rm 10}$,
B.~Abi$^{\rm 112}$,
M.~Abolins$^{\rm 88}$,
O.S.~AbouZeid$^{\rm 158}$,
H.~Abramowicz$^{\rm 153}$
% ...
\bigskip

% ...
$^{10}$ Institute of Physics, Azerbaijan Academy of Sciences, Baku, Azerbaijan\\
$^{11}$ Institut de F\'isica d'Altes Energies and Departament de F\'isica de la Universitat Aut\`onoma  de Barcelona and ICREA, Barcelona, Spain\\
% ...
$^{48}$ Fakult\"{a}t f\"{u}r Mathematik und Physik, Albert-Ludwigs-Universit\"{a}t, Freiburg i.Br., Germany\\
$^{49}$ Section de Physique, Universit\'e de Gen\`eve, Geneva, Switzerland\\
% ...
$^{88}$ Department of Physics and Astronomy, Michigan State University, East Lansing MI, United States of America\\
% ...
$^{111}$ Homer L. Dodge Department of Physics and Astronomy, University of Oklahoma, Norman OK, United States of America\\
$^{112}$ Department of Physics, Oklahoma State University, Stillwater OK, United States of America\\
% ...
$^{115}$ LAL, Universit\'e Paris-Sud and CNRS/IN2P3, Orsay, France\\
% ...
$^{153}$ Raymond and Beverly Sackler School of Physics and Astronomy, Tel Aviv University, Tel Aviv, Israel\\
% ...
$^{158}$ Department of Physics, University of Toronto, Toronto ON, Canada\\
% ...
\end{flushleft}
